\documentclass[preprint]{revtex4}
\usepackage{amssymb,amsmath}

\begin{document}

\draft

\title{Is Adaptive Landscape in Biology a Metaphor
            or a Quantitative Concept? or Both? }

\author{ Ping Ao }
\address{ Department of Mechanical Engineering and Department of Physics \\
               University of Washington, Seattle, WA 98195, USA  }
%          $^{\ddag}$Department of Physics,
%               University of Washington, Seattle, WA 98195, USA   }
% {\bf Corresponding author:}
%    $^{\%}$E-mail: aoping@u.washington.edu, Tel: 206-543-7837,
%    Fax: 206-685-8047

\date{March 23 (2007) }

%\noindent
\begin{abstract}
 Some biologists accept Wright's adaptive landscape idea,
 believing it is one of most profound concepts in evolutionary dynamics.
 Some wouldn't, believing that "the idea that there is such a
 quantity remains one of the most widely held popular misconceptions
 about evolution."
 The two groups usually have very limited communication with each other.
 The paper of Poelwijk, Kiviet, Weinreich, and Tans (Nature, 445:
 383-386(2007)) is an example.
 Sometimes such isolation can be good, because it protects budding
 ideas in a harsh environment.
 Thanks to the theoretical and experimental progress during past
 few years, time may have arrived to consider both sides seriously.
 Present letter is an attempt to bridge this gap, by pointing out
 two misrepresentations and one resulting error in Poelwijk {\it et al}.
\end{abstract}

%\pacs{PACS numbers: }

\maketitle

%\section{ Introduction }

Wright's adaptive landscape and the associated potential function
have been controversial in biology even since they were proposed
some 70 years ago \cite{wright1932}.  In his recent textbook Rice
stated that "there is no general potential function underlying
evolution. ... the idea that there is such a quantity remains one
of the most widely held popular misconceptions about evolution"
\cite{rice2004}. Despite such strong negative opinion, the
explanatory power of Wright's idea is evident and remains
attractive. So went the 2007 paper of Poelwijk {\it et al}
\cite{poelwijk2007} on the numerical construction of the landscape
for protein evolution paths. In their dash, however, Poelwijk {\it
et al} made two misrepresentations and one error.

{\ }

\noindent
{\bf Does the adaptive landscape really have a meaning
beyond its metaphoric role? } \\
By ignoring the abundant objections such as that of Rice Poelwijk
{\it et al} chose to present a consensus picture. However, apart
many intriguing "counterexamples" in literature, Rice emphatically
stressed the simple limit cycle type dynamics as a
"counterexample" for nonexistence of adaptive landscape. To my
knowledge there is no work to head-on addressing such objection,
other than those in Ref.'s \cite{ao2005,zhu2006}, which are not
cited by Poelwijk {\it et al}. Fortunately, the general conclusion
of the existence of quantitative adaptive landscape in Ref.'s
\cite{poelwijk2007,ao2005,zhu2006} all agree with each other, a
rather happy coincident.

{\ }

\noindent
{\bf What is the effect of noise?} \\
Poelwijk {\it et al} did notice that there are some situations
that the landscape seems ill-defined, such as "the situation in
which environmental fluctuations are fast relative to selection
timescales". They suggested that "recently theoretical
considerations may provide promising approaches to address these
questions more generally." This leads to their second
misrepresentation: Such general theoretical framework has already
existed since 2004 \cite{ao2005,ao2004,kat}, again not mentioned
in their work. Specifically, it was already shown in 2005 that the
existence of the Wright adaptive landscape is a general property
of Darwinian evolutionary processes \cite{ao2005}. The noise, both
genetic and environmental, is shown to be able to affect the
adaptive landscape in a profound and quantitative way
\cite{ao2005,zhu2006,ao2004,kat,zhu2004}. Here is then another and
more impressive happy coincidence.

{\ }

\noindent
{\bf "Intelligent design" vs Evolution. } \\
The exaggeration and negligence, or vision and boldness if putting
alternatively, are abundant in literature. For example, in a
recent Nature paper the results were presented in such a way it
appeared the adaptive landscape in biology started with them
\cite{oudenaarden2005}. Hence, experienced readers would ask: What
is the real fuss on such issue? Well, the reason was actually been
suggested by Poelwijk {\it et al.}: To fight the misunderstanding
on evolution we have to be extra careful.  It is known that the
advocates of "intelligent design" have the habit to ignore
overwhelming inconvenient evidence and to selectively pick up ones
"supporting" their theories. As scientists, we naturally cannot
operate on such "professional" level. Otherwise, it would not
improve the credibility of science. There is an already known
example of such rash in 1990's on the evolution of eye:
%\cite{nilsson}:
The researchers were not careful enough on assumptions and gaps in
their simulation. They led others to believe a definite conclusion
was already reached. Till these days there are still ongoing
experimental efforts to understand the evolution of eye
\cite{science2007}. With this in mind, it is worthwhile
\cite{note} to point out the works neglected by Poelwijk {\it et
al} as well as their misrepresentations.

{\ }

\noindent
{\bf Importance of drift.} \\
Poelwijk {\it et al.} did ask important questions such as "Is
neutral genetic drift essential for a new trait to emerge?"
However, it appears wrong for them to conclude that "neutral
genetic drift is not essential in case studied." Let me ignore
their rather restrictive use of drift and focus on two facts
showing in their work. First, without the drift, the wondering in
the flat part of landscape in their Fig.1c would not happen.
Second, if they perform a large sampling, they should be able to
obtain distribution determined by their adaptive landscape as
discussed in Ref.\cite{ao2005}, even for single adaptive peak.
There would be no such distribution if there were no drift. Thus,
drift cannot be ignored.

There are two additional reasons that their negative conclusion on
drift is not correct. There is a so-called fundamental theorem of
natural selection \cite{fisher1930} generalized in
Ref.\cite{ao2005}: The rate to approach to the peak is
proportional to the drift or variation. Zero drift would imply the
impossibility to reach the adaptive peak! Furthermore, as
discussed by Wright \cite{wright1932,ao2005}, the drift would make
the transition from one peak to another, such depicted in their
Fig.1b. A report on the quantitative agreement between
experimental and theoretical studies of such behaviors in the
phage lambda may serve as a nontrivial example of such transition
\cite{zhu2004}. The real problem seems to be time scales. It is
likely that in their numerical experiment Poelwijk {\it et al} had
not run their time long enough to reveal such features.

I do not believe Poelwijk {\it et al} had deliberated ignored
relevant prior works: Once the two misrepresentations and one
error are corrected, their main conclusion of the existence of
adaptive landscape becomes stronger.

{\ }

\end{document}